# Giant enhancement in spin-to-charge conversion in $Bi_2Se_3$/NiFe heterostructure via interface engineering

Sourav Rajat Subhra Maitra[1], Poulami Manna[2], Ravi Prakash Singh[2], Chandrasekhar Murapaka[3], Arabinda Haldar[1,*]

[1]*Department of Physics, Indian Institute of Technology Hyderabad, Kandi-502285, Telangana, India*

[2]*Department of Physics, Indian Institute of Science Education and Research Bhopal, Bhopal, 462066, India*

[3]*Department of Material Science and Metallurgical Engineering, Indian Institute of Technology Hyderabad, Kandi-502285, Telangana, India*

*Corresponding author: Electronic address: arabinda@phy.iith.ac.in

## ABSTRACT

Topological insulators provide a promising platform for spintronic applications owing to their spin-momentum-locked surface states and efficient spin-to-charge conversion. Among these, $Bi_2Se_3$ has been the subject of intensive investigation due to its large bulk bandgap and single Dirac cone band structure. However, spin-to-charge conversion strongly depends on the quality of the topological insulator/ferromagnet interface. Here, we investigate spin transport and spin-to-charge conversion in sputter-deposited $Bi_2Se_3$/Ti/NiFe heterostructures at room temperature. The topological insulator layer is deposited on a CMOS-compatible silicon substrate. Low-temperature magnetoresistance measurements are conducted to establish the existence of a surface conducting channel in the deposited topological insulator layer. Pure spin current was injected into the $Bi_2Se_3$ layer through the Titanium spacer layer via spin pumping induced by the spin precession in microwave-driven ferromagnetic resonance of the ferromagnetic film. Spin pumping studies are carried out by varying the thickness of the $Bi_2Se_3$ layer. The $Bi_2Se_3$ thickness dependence of the Gilbert damping reveals a pronounced ~55% enhancement at a thickness of 4 nm, consistent with the hybridization of the top and bottom surface states of the topological insulator layer. The spin Hall angle – a parameter that quantifies the spin-to-charge conversion efficiency – exhibits an approximately one-order-of-magnitude enhancement upon insertion of the Ti spacer compared with $Bi_2Se_3$/NiFe heterostructures without the spacer. The significant enhancement of the spin Hall angle is attributed to Ti, which inhibits interdiffusion between the $Bi_2Se_3$ and NiFe layers, thereby protecting the topological surface states. Our findings highlight titanium spacers' role in topological spintronics applications.

The overall performance of a spin-orbit torque-based spintronic device is fundamentally governed by the spin-to-charge interconversion efficiency of the constituent spin Hall material. In this context, topological insulators (TIs) play a critical role with giant spin-to-charge conversion efficiency due to their spin-polarised topological surface states (TSSs) [1–10]. The existence of TSSs in TI thin films is observed directly using angle-resolved photoemission spectroscopy (ARPES) and low-temperature magnetoresistance (MR) measurements [11–14]. Additionally, the spin-to-charge conversion is also observed in TI/ferromagnet (FM) heterostructures using angle-dependent magnetoresistance (ADMR), spin-torque ferromagnetic resonance (ST-FMR) measurements and ferromagnetic resonance (FMR) based spin pumping studies [3–6,15–18]. The quantitative parameter for spin-to-charge conversion is the spin Hall angle (SHA), defined as the ratio of the charge current density ($J_c$) to the spin current density ($J_s$), generated by the FM layer in the TI/FM heterostructure. Giant SHA observed in TIs is primarily due to the large spin-orbit coupling (SOC) and the perpendicular locking of spin polarisation with their momentum [3,7,9,10,16]. The uniform spin oscillation generated in the ferromagnetic layer propagates through the TI/FM interface and is scattered in the TI layer. This scattering results in a transverse charge current generated perpendicular to the spin current, which is known as the inverse spin Hall effect (ISHE) [17,19]. Additionally, scattering occurs in the spin-polarised TSS through the interfacial Rashba effect [19–21]. This results in an additional charge current due to the inverse Rashba-Edelstein effect (IREE). These two results show high spin-to-charge conversion efficiency in the TI layer. Among various TIs, $Bi_2Se_3$, BiSb, and $Bi_2Te_3$ are found to be efficient TIs for spin-to-charge conversion [3,5,6,22–24].

To date, high-quality crystalline TI layers have been epitaxially grown on substrates such as sapphire, STO and MgO using pulsed laser deposition (PLD) or molecular beam epitaxy (MBE) [15,16,25,26]. The techniques are not compatible with conventional semiconductor process flows and thus cannot be scaled. Further, the melting point of the TI materials poses challenges for the fabrication of sharp interfaces in TI/FM heterostructures[27]. The TI layer, due to its low melting point, tends to interdiffuse with the FM layer, leading to the suppression or disintegration of the TI-layer TSSs [28,29]. In recent years, the use of spacer layers has been explored to block the interdiffusion, thereby increasing spin-to-charge conversion. The SHA in TI/FM heterostructures with different spacer layers, such as Cu, $Cr_2O_3$, Ag, Al, Te and MgO, has been studied[30–33]. Additionally, the effect of antiferromagnetic spacer layers on the SHA has been reported[34,35]. Recently, Titanium (Ti) spacer layers have enhanced SOT efficiency in BiSb/NiFe heterostructures by suppressing interfacial interdiffusion while facilitating efficient spin-current transmission owing to Ti's relatively low SOC[29]. In this context, $Bi_2Se_3$ is an excellent choice as a TI material due to its single Dirac point in the band structure and larger band gap (~ 0.3 eV), which makes the TSS more stable at room temperature than other TI materials. However, the investigation of spin-to-charge conversion in $Bi_2Se_3$/FM heterostructures is comparatively more challenging than in other TI/FM systems such as BiSb and $Bi_2Te_3$, primarily due to the high diffusivity of Se atoms, which facilitates intermixing at the interface with the ferromagnetic layer.

Here, we report an investigation of robust spin-to-charge current conversion in $Bi_2Se_3$/Ti/NiFe heterostructure at room temperature. The $Bi_2Se_3$ layer is deposited by magnetron sputtering onto silicon substrates, demonstrating a deposition route compatible with scalable device fabrication. The variation in the Gilbert damping parameter with changes in the $Bi_2Se_3$ layer thickness is measured using ferromagnetic resonance (FMR) spectroscopy. The SHA is quantified from ISHE measurements in $Bi_2Se_3$/Ti/NiFe heterostructures and compared with a corresponding $Bi_2Se_3$/NiFe control series without the Ti spacer. This comparison enables us to elucidate the role of the Ti interlayer in modifying the interfacial spin transport and spin-to-charge conversion efficiency.

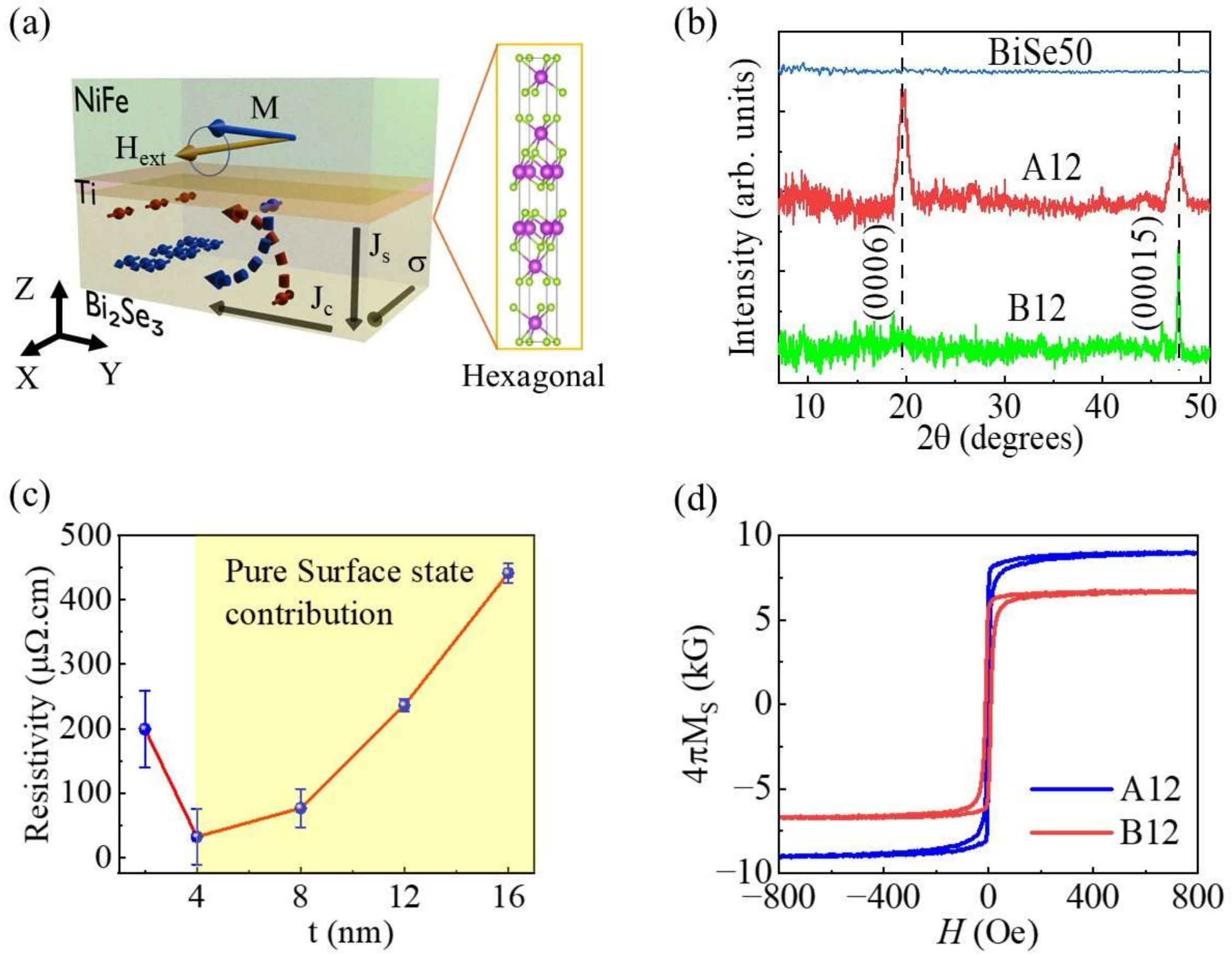

Figure 1: (a) Schematic diagram of spin-to-charge conversion in the heterostructure, with the crystal structure of $Bi_2Se_3$, (b) the XRD pattern (θ-2θ scan) of A12: $Bi_2Se_3$ (12)/Ti (3)/NiFe (10), B12: $Bi_2Se_3$ (12)/NiFe (10) and BiSe50: as-deposited $Bi_2Se_3$ (50) samples, respectively. The vertical dashed lines represent the diffraction peaks corresponding to different c-axis-oriented planes, (c) magnetization reversal curve of the samples A12: $Bi_2Se_3$ (12)/Ti (3)/NiFe (10) and B12: $Bi_2Se_3$ (12)/NiFe (10), (d) variation in the resistivity of crystalline $Bi_2Se_3$ thin films with respect to their thicknesses.

Two sets of samples were fabricated, named Series A: $Bi_2Se_3$ (t)/Ti (3)/NiFe (10) and Series B: $Bi_2Se_3$ (t)/NiFe (10), where t = 0, 2, 4, 8, 12, 16 nm on silicon (100) substrates using magnetron sputtering. An as-deposited $Bi_2Se_3$ (50) thin film, named BiSe50, is fabricated to compare the crystallinity with the annealed samples. Furthermore, a Hall bar structure of crystalline $Bi_2Se_3$ (50) is fabricated to investigate the Two-Dimensional Weak Anti-Localization (2D-WAL) effect in the TI thin films (Supplementary Materials, S1). The schematic diagram of the heterostructure of series A is shown in Fig. 1(a).

FMR measurements are performed using a lock-in-based spectrometer (NanOsc Instruments), while the ISHE voltage is detected using a Keithley 2182A nanovoltmeter (Supplementary Materials, S2). The ISHE phenomenon in the heterostructure is shown in Fig. 1(a).

Figure 1(b) presents the out-of-plane XRD θ-2θ patterns of samples A12, B12, and BiSe50. The observed reflections at (0 0 0 6) and (0 0 0 15) confirm polycrystalline $Bi_2Se_3$ with a preferential (0 0 0 n) texture, indicating that a substantial fraction of the c-axis is oriented in-plane. The enhanced diffraction intensity of A12 relative to B12 suggests improved crystallinity in the presence of the spacer layer. In contrast, the as-deposited BiSe50 film exhibits no discernible diffraction peaks, indicating a predominantly amorphous structure and, consequently, suppressed TSS. Thus, to improve crystallinity, the $Bi_2Se_3$ films were annealed at 300 °C for 1 h. The resistivity of the $Bi_2Se_3$ thin films with respect to their thicknesses is shown in Fig. 1(c). The increase in resistivity indicated decoupling of the top and bottom surface states, ensuring a pure surface-state contribution beyond 4 nm.

Furthermore, to assess the elemental composition of $Bi_2Se_3$ thin films, we performed energy-dispersive X-ray spectroscopy (EDS) using scanning electron microscopy (SEM). The atomic percentages of Se and Bi have been found to be 63.21% and 36.79%, respectively, which is almost a 3:2 composition ratio. This ensures the proper growth and composition of $Bi_2Se_3$ thin films (Supplementary Materials, S3). The composition is essential for pure surface states.

Moreover, the magnetization reversal curve is obtained using a vibrating sample magnetometer (VSM) to determine the saturation magnetization of the samples. The magnetization curves of the samples A12 and B12 are shown in Fig. 1(d). Notably, the saturation magnetization of sample B12 is almost 35% lower than that of sample A12, which can be attributed to interfacial intermixing between Fe atoms and the $Bi_2Se_3$ layer. The saturation magnetization of the samples A0 and B0 are evaluated using VSM measurements, which are used to determine the SHA of the samples (Supplementary Materials, S4).

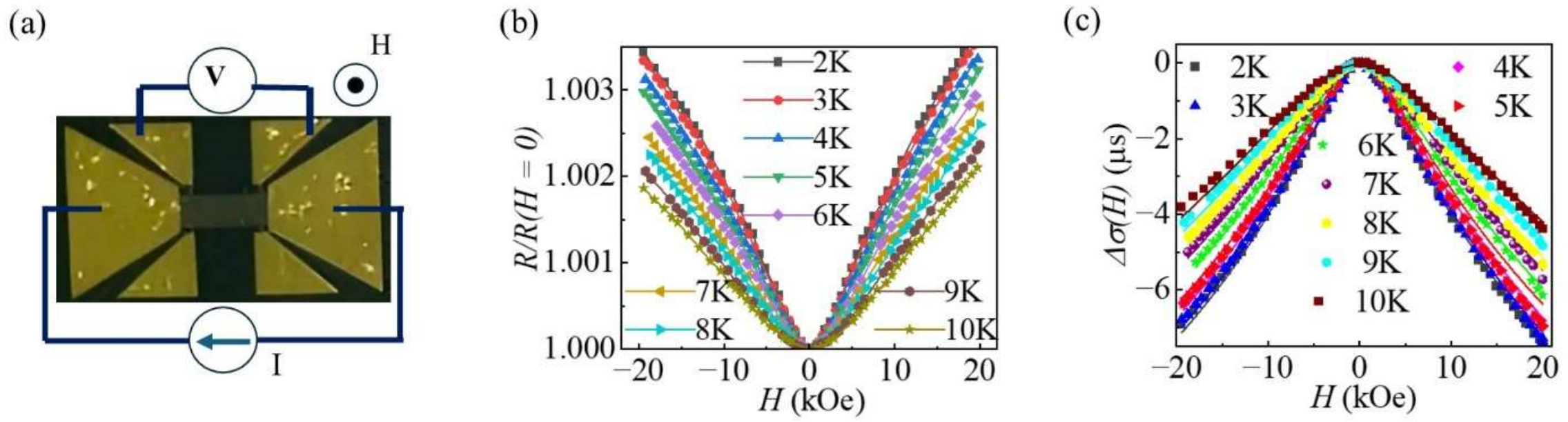

Figure 2: (a) An optical micrograph of the Hall bar structure of dimensions 200 μm × 100 μm with the schematic of the magnetoresistance measurement setup at temperatures T = 2K to 10K. Here, the magnetic field H is directed outward from the plane of the paper, (b) Variation in the relative magnetoresistance of 50 nm crystalline $Bi_2Se_3$ Hall bar at temperatures T = 2K to 10K, (c) Variation of the magnetoconductivity of 50 nm crystalline $Bi_2Se_3$ Hall bar at temperatures T = 2K to 10K.

To further substantiate the existence of topological surface states (TSS) in the $Bi_2Se_3$ thin films, low-temperature MR measurements were performed on a 50 nm crystalline $Bi_2Se_3$ Hall bar device of dimensions 200 μm × 100 μm using a physical property measurement system (PPMS). The temperature was varied from 2 K to 10 K in 1 K steps. At each temperature, the out-of-plane magnetic field was swept from +20,000 Oe to −20,000 Oe while a constant current of 1 mA was applied to record the resistance as a function of magnetic field. The optical micrograph of the Hall bar geometry along with the measurement configuration is shown in Fig. 2(a). The relative MR, presented in Fig. 2(b), exhibits a positive dependence on the applied out-of-plane magnetic field, consistent with WAL behaviour arising from strong SOC and symmetry considerations in TIs. The corresponding variation of the magnetoconductance as a function of magnetic field is calculated using the following equation:

$$\Delta\sigma(H) = \frac{L}{W(R(H)-R(0))} \tag{1}$$

where $H$, $R(H)$, and $R(0)$ represent the applied magnetic field, the resistance at the field $H$, and the zero-field resistance, respectively, while $L$ and $W$ denote the length and width of the Hall bar device. Figure 2(c) presents the variation of magnetoconductance as a function of the applied magnetic field. To quantitatively analyse the WAL behaviour, the magnetoconductance data were fitted using the Hikami–Larkin–Nagaoka (HLN) model, expressed as follows[36]:

$$\Delta\sigma(H) = -\frac{\alpha e^2}{2\pi^2\hbar}\left[\ln\left(\frac{\hbar}{4Hel_\varphi^2}\right) + \Psi\left(\frac{1}{2} + \frac{\hbar}{4Hel_\varphi^2}\right)\right] \tag{2}$$

where $\alpha$, $l_{\phi}$, and $H$ represent the prefactor, phase coherence length, and applied magnetic field, respectively. The application of an external magnetic field breaks the time-reversal symmetry of the TSS, giving rise to the WAL effect in the $Bi_2Se_3$ thin films. Consequently, the constructive quantum interference responsible for enhanced conductivity at zero field is suppressed, leading to a decrease in magnetoconductivity with increasing magnetic field[13,37]. The extracted value of the pre-factor, $\alpha \approx -0.76$, indicates the presence of at least one coherent surface conduction channel in the $Bi_2Se_3$ thin films.

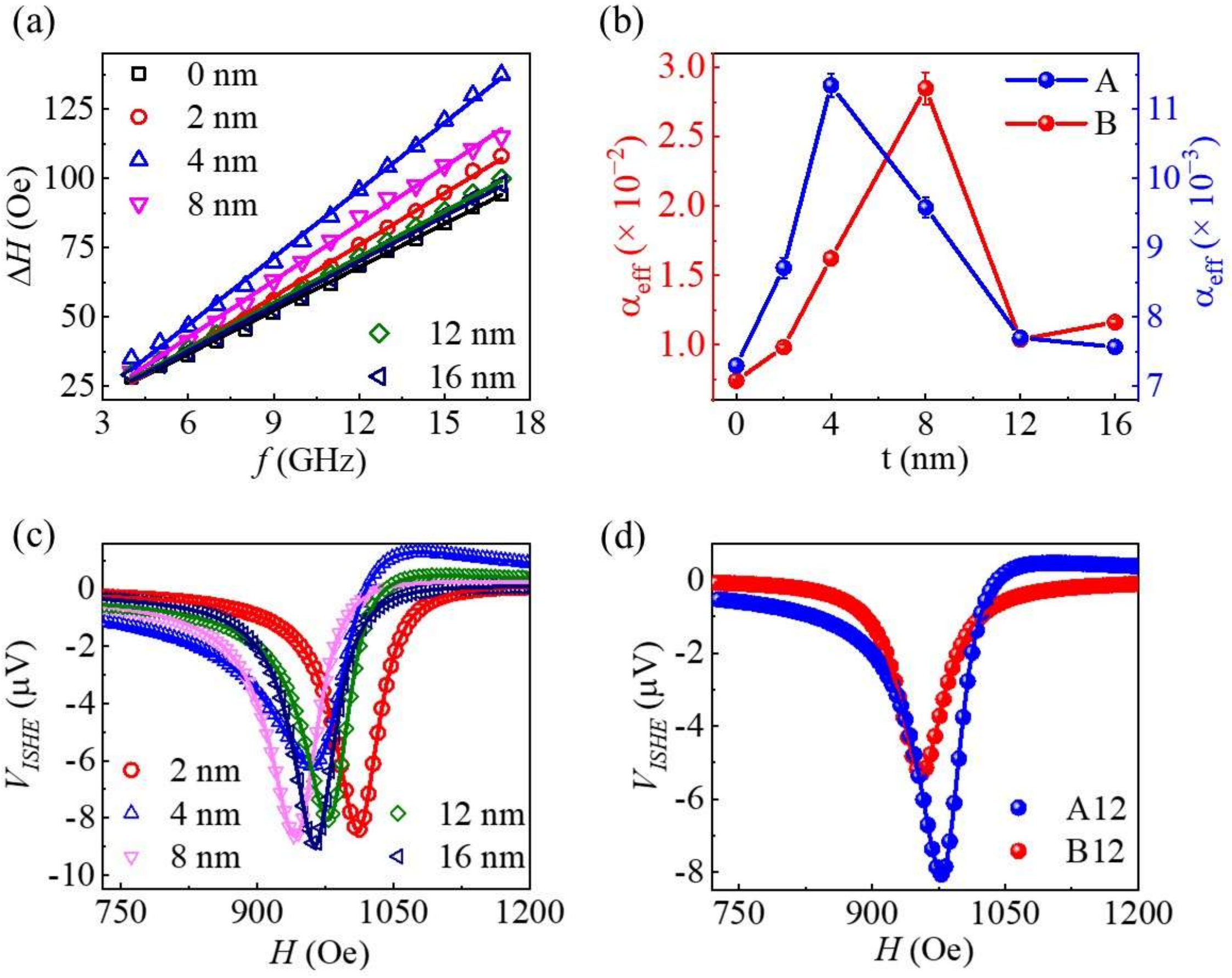

Figure 3: (a) Linewidth ($\Delta H$) vs. Frequency ($f$) plot in the sample $Bi_2Se_3$ (t)/Ti (3)/NiFe (10), where t = 0, 2, 4, 8, 12, 16 nm, (b) Variation of the effective Gilbert damping parameter ($\alpha_{eff}$) with the thickness ($t$) of the $Bi_2Se_3$ layer in the series-A: $Bi_2Se_3$ (t)/Ti (3)/NiFe (10) and the series-B: $Bi_2Se_3$ (t)/NiFe (10), respectively, (c) ISHE voltage at *rf* excitation at 9 GHz in the sample $Bi_2Se_3$ (t)/Ti (3)/NiFe (10), where t = 2, 4, 8, 12, 16 nm, (d) ISHE voltage drop at *rf* excitation frequency of 9 GHz of the samples A12: $Bi_2Se_3$ (12)/Ti (3)/NiFe (10) and B12: $Bi_2Se_3$ (12)/NiFe (10), respectively.

The FMR spectroscopy in the samples $Bi_2Se_3$ (t)/Ti (3)/NiFe (10), where t = 0, 2, 4, 8, 12, 16 nm is performed to evaluate the Gilbert damping parameter, which signifies the spin pumping from the FM layer to the TI layer (Supplementary Material, S5). The FMR spectra for the samples $Bi_2Se_3$ (t)/Ti (3)/NiFe (10) are recorded in the frequency range of 4-17 GHz, and the extracted linewidth $\Delta H$ of the spectra are plotted with respect to the frequency $f$, which is shown in Fig. 3(a). The dependence of $\Delta H$ can be fitted with $f$ using the equation[38],

$$\Delta H = \Delta H_0 + \frac{4\pi}{\gamma} f \tag{3}$$

where, $\alpha$ and $\Delta H_0$ are the Gilbert damping constant and inhomogeneous linewidth broadening, respectively. Furthermore, we extracted the Gilbert damping parameter for each $Bi_2Se_3$ (t)/Ti (3)/NiFe (10) sample from the slopes of $\Delta H$ as a function of $f$, as shown in Fig. 3(a). The value of the effective Gilbert damping parameter can be written as, $\alpha_{eff} = \alpha_0 + \Delta\alpha$, where $\alpha_0$ is the intrinsic damping of the 10 nm NiFe sample, which is broadened by $\Delta\alpha$ because of the spin pumping to the adjacent $Bi_2Se_3$ layer. The precessing magnetization of the NiFe layer is driven by the *rf* magnetic field, inducing spin pumping into the TI layer. The pumped angular momentum is subsequently dissipated through scattering by the TSS and bulk SOC. The resulting spin loss at the NiFe/$Bi_2Se_3$ interface enhances the effective Gilbert damping, as shown in Fig. 3(b). The damping increases with $Bi_2Se_3$ thickness up to 4 nm, consistent with hybridization between the top and bottom surface states. Beyond 4 nm, these states decouple, restricting spin scattering predominantly to the interfacial surface state and reducing the damping. This behaviour indicates dominant surface-state-mediated spin scattering for thicknesses above 4 nm. Similar observations were made by Xue *et al.* in Fe/$Bi_2Te_3$ heterostructure, where hybridization of the surface states was shown using spin-resolved ARPES measurements [39]. This observation is consistent with the resistivity behaviour presented in Fig. 1(c).

To verify the effective saturation magnetization in the samples $Bi_2Se_3$ (t)/Ti (3)/NiFe (10), where t = 0, 2, 4, 8, 12, 16 nm, we analysed the variation of the resonance magnetic field with respect to the resonance frequency (*f*) with the Kittel equation (Supplementary Material, S5). Moreover, we also evaluated spin-mixing conductance and subsequently, the spin current in the samples $Bi_2Se_3$ (t)/Ti (3)/NiFe (10), to investigate the spin-to-charge conversion (Supplementary Materials, S5).

Furthermore, we explored the spin-to-charge conversion capability of these samples by measuring the ISHE voltages. The observed reversal in the voltage polarity at equal magnitudes of the applied magnetic field in sample A12 confirms the characteristic signature of the ISHE in the sample (Supplementary Materials, S5). Further, to quantify the voltage contribution from the ISHE among all other spin rectification effects in our samples, the measured voltage signal was fitted to the Lorentzian equation with symmetric and anti-symmetric contributions[40],

$$V_{ISHE} = V_{sym}\left[\frac{(\Delta H)^2}{(H-H_{res})^2+(\Delta H)^2}\right] + V_{asym}\left[\frac{2\Delta H(H-H_{res})}{(H-H_{res})^2+(\Delta H)^2}\right] \quad (4)$$

where, $V_{sym}$ and $V_{asym}$ are the symmetric and anti-symmetric parts of the voltage drop, respectively. The symmetric component predominantly reflects the spin-pumping contribution to the ISHE signal, whereas the anti-symmetric component arises from the anomalous Hall effect (AHE). Fig. 3(c) shows the ISHE voltage drop for $Bi_2Se_3$ (t)/Ti (3)/NiFe (10) samples with t = 2, 4, 8, 12, and 16 nm, at 9 GHz *rf* excitation. It clearly shows that the voltage peak decreases sharply from 2 nm to 4 nm of $Bi_2Se_3$ layer thickness, due to the decoupling of the surface states. It then nearly saturates at approximately 8 nm, indicating that the bottom-surface state is the dominant scattering mechanism.

Furthermore, to evaluate the spin-pumping contribution to the ISHE voltage, we measured the ISHE voltage as a function of the angle $\varphi$ of the applied DC magnetic field for the A12 sample (Supplementary Materials, S5). The value of $V_{sp}$ obtained for A12 is −15.3232 μV.

For comparison, FMR and ISHE measurements were performed on $Bi_2Se_3$ (t)/NiFe (10) control samples with t = 0, 2, 4, 8, 12, and 16 nm (Supplementary Materials, S6). Figure 3(b) shows the variation of the effective Gilbert damping parameter ($\alpha_{eff}$) with the thickness of the TI layer in the samples $Bi_2Se_3$ (t)/NiFe (10). It shows a sharp peak at t = 8 nm, the same as in sample series A, which indicates decoupling of the top and bottom surface states beyond t = 8 nm. Notably, the peak in the variation of the Gilbert damping parameter shifts to a TI layer thickness of 8 nm. This shift can be attributed to a reduction in the effective TI thickness arising from intermixing at the TI/FM interface. Moreover, to estimate the spin-to-charge conversion efficiency in the samples $Bi_2Se_3$ (t)/NiFe (10), where t = 0, 2, 4, 8, 12, 16 nm, we performed ISHE measurements (Supplementary Materials, S6). Figure 3(d) shows the ISHE voltage drop in the sample B12: $Bi_2Se_3$ (12)/NiFe (10), which clearly shows a decrease in the ISHE voltage compared to sample A12. Furthermore, to estimate the spin pumping voltage $V_{sp}$, we have performed angular ISHE measurements on the sample $Bi_2Se_3$ (12)/NiFe (10) (Supplementary Materials, S6). The value of $V_{sp}$ obtained for the sample B12 is –9.7491 μV, which is less than that of A12. This is evidence of increased spin-to-charge conversion by introducing a Ti spacer layer.

Subsequently, we have calculated the SHA for both samples, A12 and B12. The spin pumping voltage $V_{sp}$ and the spin current $J_S$ of any FM/HM heterostructure can be related to the spin current $J_S$ using the following equation[41],

$$V_{SP} = \left( \frac{1}{\frac{t_{NiFe}}{\rho_{NiFe}} + \frac{t_{Ti}}{\rho_{Ti}} + \frac{t}{\rho}} \right) w \theta_{SH} \lambda_{sd} \tanh\left(\frac{t}{2\lambda_{sd}}\right) |J_s| \tag{5}$$

$$V_{SP} = \left( \frac{1}{\frac{t_{NiFe}}{\rho_{NiFe}} + \frac{t}{\rho}} \right) w \theta_{SH} \lambda_{sd} \tanh\left(\frac{t}{2\lambda_{sd}}\right) |J_s| \tag{6}$$

where, $\rho_{Py}$, $\rho$, $t$, $t_{Ti}$, $\rho_{Ti}$, $w$, $\lambda_{sd}$ and $\theta_{SH}$ are the resistivity of the FM layer, resistivity of the $Bi_2Se_3$ layer, the thickness of the $Bi_2Se_3$ layer, the thickness of the Ti layer, which is 3 nm, resistivity of the Ti layer, signal line width (= 200 μm), spin diffusion length of the $Bi_2Se_3$ layer (~ 6.2 nm), and the SHA, respectively. The calculated SHA for the samples A12 and B12 using Eq. (11) and Eq. (12), respectively, are shown in the following Table 1.

Table 1: Values of $4\pi M_s$, $J_S$, $V_{SP}$ and SHA for the samples $Bi_2Se_3$ (12)/Ti (3)/NiFe (10) and $Bi_2Se_3$ (12)/NiFe (10)

| Sample | $4\pi M_s$ (G) | $J_s$ (erg/cm$^2$) | $V_{SP}$ (μV) | $\theta_{SH}$ |
|---|---|---|---|---|
| A12: $Bi_2Se_3$ (12)/Ti (3)/NiFe (10) | 9007.4 ± 79.6 | 466288440 | 15.32 ± 0.07 | 1.36 ± 0.03 |
| B12: $Bi_2Se_3$ (12)/NiFe (10) | 6657.9 ± 94.8 | 1848866190 | 9.75 ± 0.01 | 0.14 ± 0.06 |

Notably, the spin current in sample B12 is higher than that in sample A12. The spin pumping generated at the FM layer of the sample A12 must pass through two interfaces: one is NiFe/Ti, and the other is Ti/$Bi_2Se_3$, which causes a significant loss in the spin current, though the Ti spacer layer is not contributing to the spin-to-charge conversion due to its low SOC[42]. Nevertheless, the $\theta_{SH}$ of the A12 sample is 1.36 ± 0.03, significantly exceeding that of the B12 sample, which is 0.14 ± 0.06. This indicates that sample A12 generates a substantially larger spin-pumping-induced voltage even at a lower spin current compared to sample B12. This is a remarkable increase in spin-to-charge conversion resulting from the insertion of the Ti spacer layer. Mellnik *et al.* reported a SHA of 2.0–3.5 for $Bi_2Se_3$ thin films grown by MBE on sapphire substrates [3]. In contrast, our results demonstrate a nearly tenfold enhancement in the SHA upon the introduction of a spacer layer. Furthermore, our SHA value is significantly higher than that for samples grown by magnetron sputtering on silicon substrates[16,17]. The reason for this enhancement is the blocking of interdiffusion between the FM and TI layers. The Fe atoms from the NiFe layer tend to diffuse with the Se atoms of the $Bi_2Se_3$ layer, forming $FeSe_x$[28]. This interdiffusion resists the pumped spin current from reaching the $Bi_2Se_3$ layer. Moreover, the magnetic impurities erode the topological surface states of the $Bi_2Se_3$ layer by breaking its $\pi$ Berry phase [13,14]. This impurity introduces weak localisation, which is the main cause of the absence of surface states in the $Bi_2Se_3$ (2-16)/NiFe (10) samples as reported by H. He et al[13,37].

In summary, we demonstrate enhanced spin-to-charge conversion in CMOS-compatible sputter-deposited $Bi_2Se_3$/NiFe heterostructures through Ti spacer insertion. FMR measurements reveal a pronounced thickness dependence of the Gilbert damping, with surface-state-mediated spin transport emerging for $Bi_2Se_3$ thicknesses above 4 nm. The SHA increases from $0.14 \pm 0.06$ without the spacer to $1.36 \pm 0.03$ with the Ti spacer, corresponding to nearly a tenfold enhancement. This improvement is attributed to reduced interfacial interdiffusion and improved preservation of the $Bi_2Se_3$ surface states. These results demonstrate an effective interface-engineering approach for enhancing spin-to-charge conversion in topological-insulator-based heterostructures.

See the Supplementary Materials for detailed sample fabrication and deposition processes, FMR and ISHE measurement techniques, energy-dispersive X-ray (EDS) spectroscopy, VSM measurements and detailed FMR and ISHE measurements and analysis in the samples.

A.H. would like to thank the funding from the ANRF (ANRF/ARG/2025/009623/PS). C.M. would like to acknowledge funding from SERB-Core Research Grant CRG/2022/005472. S.R.S.M. would like to thank the scholarship provided by the Ministry of Education (MoE), Government of India.

## I. DATA AVAILABILITY

The data that support the findings of this study are available from the corresponding author upon reasonable request.

## II. CONFLICT OF INTEREST

The authors have no conflicts of interest to disclose.

## III. AUTHOR CONTRIBUTIONS

**Sourav Rajat Subhra Maitra:** Conceptualization (equal); Data curation (lead); Formal Analysis (lead); Investigation (equal); Methodology (lead); Writing – original draft (lead). **Poulami Manna:** Data curation. **Ravi Prakash Singh:** Investigation (equal); Writing-review & editing (equal). **Chandrasekhar Murapaka:** Conceptualization (equal); Funding acquisition (equal); Investigation (equal); Resources (equal); Supervision (equal); Writing-review & editing (equal). **Arabinda Haldar:** Conceptualization (equal); Funding acquisition (equal); Investigation (equal); Resources (equal); Supervision (equal); Writing-review & editing (equal).

## REFERENCES

[1] L. Fu, C.L. Kane, and E.J. Mele, "Topological insulators in three dimensions," Phys. Rev. Lett. **98**(10), 106803 (2007).

[2] D. Hsieh, D. Qian, L. Wray, Y. Xia, Y.S. Hor, R.J. Cava, and M.Z. Hasan, "A topological Dirac insulator in a quantum spin Hall phase," Nature **452**(7190), 970–974 (2008).

[3] A.R. Mellnik, J.S. Lee, A. Richardella, J.L. Grab, P.J. Mintun, M.H. Fischer, A. Vaezi, A. Manchon, E.A. Kim, N. Samarth, and D.C. Ralph, "Spin-transfer torque generated by a topological insulator," Nature **511**(7510), 449–451 (2014).

[4] Y. Fan, P. Upadhyaya, X. Kou, M. Lang, S. Takei, Z. Wang, J. Tang, L. He, L. Te Chang, M. Montazeri, G. Yu, W. Jiang, T. Nie, R.N. Schwartz, Y. Tserkovnyak, and K.L. Wang, "Magnetization switching through giant spin-orbit torque in a magnetically doped topological insulator heterostructure," Nat. Mater. **13**(7), 699–704 (2014).

[5] K. Kondou, R. Yoshimi, A. Tsukazaki, Y. Fukuma, J. Matsuno, K.S. Takahashi, M. Kawasaki, Y. Tokura, and Y. Otani, "Fermi-level-dependent charge-to-spin current conversion by Dirac surface states of topological insulators," Nat. Phys. **12**(11), 1027–1031 (2016).

[6] Y. Wang, P. Deorani, K. Banerjee, N. Koirala, M. Brahlek, S. Oh, and H. Yang, "Topological surface states originated spin-orbit torques in $Bi_2Se_3$," Phys. Rev. Lett. **114**(25), 257202 (2015).

[7] M.Z. Hasan, and C.L. Kane, "Colloquium: Topological insulators," Rev. Mod. Phys. **82**(4), 3045–3067 (2010).

[8] H. Zhang, C.X. Liu, X.L. Qi, X. Dai, Z. Fang, and S.C. Zhang, "Topological insulators in $Bi_2Se_3$, $Bi_2Te_3$ and $Sb_2Te_3$ with a single Dirac cone on the surface," Nat. Phys. **5**(6), 438–442 (2009).

[9] C.H. Li, O.M.J. Van't Erve, J.T. Robinson, Y. Liu, L. Li, and B.T. Jonker, "Electrical detection of charge-current-induced spin polarization due to spin-momentum locking in $Bi_2Se_3$," Nat. Nanotechnol. **9**(3), 218–224 (2014).

[10] Y. Ando, T. Hamasaki, T. Kurokawa, K. Ichiba, F. Yang, M. Novak, S. Sasaki, K. Segawa, Y. Ando, and M. Shiraishi, "Electrical detection of the spin polarization due to charge flow in the surface state of the topological insulator $Bi_{1.5}Sb_{0.5}Te_{1.7}Se_{1.3}$," Nano Lett. **14**(11), 6226–6230 (2014).

[11] Y.L. Chen, J.G. Analytis, J.-H. Chu, Z.K. Liu, S.-K. Mo, X.L. Qi, H.J. Zhang, D.H. Lu, X. Dai, Z. Fang, S.C. Zhang, I.R. Fisher, Z. Hussain, and Z.-X. Shen, "Experimental Realization of a Three-Dimensional Topological Insulator, $Bi_2Te_3$," Science (1979). **325**, 178–181 (2009).

[12] Z.H. Pan, E. Vescovo, A. V. Fedorov, D. Gardner, Y.S. Lee, S. Chu, G.D. Gu, and T. Valla, "Electronic structure of the topological insulator $Bi_2Se_3$ using angle-resolved photoemission spectroscopy: Evidence for a nearly full surface spin polarization," Phys. Rev. Lett. **106**(25), 257004 (2011).

[13] H.T. He, G. Wang, T. Zhang, I.K. Sou, G.K.L. Wong, J.N. Wang, H.Z. Lu, S.Q. Shen, and F.C. Zhang, "Impurity effect on weak antilocalization in the topological insulator $Bi_2Te_3$," Phys. Rev. Lett. **106**(16), 166805 (2011).

[14] J.J. Cha, M. Claassen, D. Kong, S.S. Hong, K.J. Koski, X.L. Qi, and Y. Cui, "Effects of magnetic doping on weak antilocalization in narrow $Bi_2Se_3$ nanoribbons," Nano Lett. **12**(8), 4355–4359 (2012).

[15] X. Zhang, C.H. Li, J. Moon, S. Leontsev, M.R. Page, B.T. Jonker, and O. Van 'T Erve, "Interplay of spin current and magnetization in a topological-insulator/magnetic-insulator bilayer structure," Phys. Rev. Mater. **6**(7), 074203 (2022).

[16] M. Jamali, J.S. Lee, J.S. Jeong, F. Mahfouzi, Y. Lv, Z. Zhao, B.K. Nikolić, K.A. Mkhoyan, N. Samarth, and J.P. Wang, "Giant Spin Pumping and Inverse Spin Hall Effect in the Presence of Surface and Bulk Spin-Orbit Coupling of Topological Insulator $Bi_2Se_3$," Nano Lett. **15**(10), 7126–7132 (2015).

[17] P. Deorani, J. Son, K. Banerjee, N. Koirala, M. Brahlek, S. Oh, and H. Yang, "Observation of inverse spin hall effect in bismuth selenide," Phys. Rev. B Condens. Matter Mater. Phys. **90**(9), 094403 (2014).

[18] J.C. Rojas-Sánchez, N. Reyren, P. Laczkowski, W. Savero, J.P. Attané, C. Deranlot, M. Jamet, J.M. George, L. Vila, and H. Jaffrès, "Spin pumping and inverse spin hall effect in platinum: The essential role of spin-memory loss at metallic interfaces," Phys. Rev. Lett. **112**(10), 106602 (2014).

[19] W. Luo, W.Y. Deng, H. Geng, M.N. Chen, R. Shen, L. Sheng, and D.Y. Xing, "Perfect inverse spin Hall effect and inverse Edelstein effect due to helical spin-momentum locking in topological surface states," Phys. Rev. B **93**(11), 115118 (2016).

[20] T. Suzuki, S. Fukami, N. Ishiwata, M. Yamanouchi, S. Ikeda, N. Kasai, and H. Ohno, "Current-induced effective field in perpendicularly magnetized Ta/CoFeB/MgO wire," Appl. Phys. Lett. **98**(14), 142505 (2011).

[21] A. Manchon, H.C. Koo, J. Nitta, S.M. Frolov, and R.A. Duine, "New perspectives for Rashba spin-orbit coupling," Nat. Mater. **14**(9), 871–882 (2015).

[22] N.H.D. Khang, Y. Ueda, and P.N. Hai, "A conductive topological insulator with large spin Hall effect for ultralow power spin–orbit torque switching," Nat. Mater. **17**(9), 808–813 (2018).

[23] H.H. Huy, J. Sasaki, N.H.D. Khang, S. Namba, P.N. Hai, Q. Le, B. York, C. Hwang, X. Liu, M. Gribelyuk, X. Xu, S. Le, M. Ho, and H. Takano, "Large inverse spin Hall effect in BiSb topological insulator for 4 Tb/in2magnetic recording technology," Appl. Phys. Lett. **122**(5), 052401 (2023).

[24] T. Shirokura, and P.N. Hai, "High temperature spin Hall effect in topological insulator," Appl. Phys. Lett. **122**(23), 232404 (2023).

[25] J. Han, A. Richardella, S.A. Siddiqui, J. Finley, N. Samarth, and L. Liu, "Room-Temperature Spin-Orbit Torque Switching Induced by a Topological Insulator," Phys. Rev. Lett. **119**(7), 077702 (2017).

[26] D. Zhu, Y. Wang, S. Shi, K.L. Teo, Y. Wu, and H. Yang, "Highly efficient charge-to-spin conversion from in situ $Bi_2Se_3$/Fe heterostructures," Appl. Phys. Lett. **118**(6), 062403 (2021).

[27] Y.R. Sapkota, and D. Mazumdar, "Influence of post-deposition annealing on the transport properties of sputtered $Bi_2Se_3$ thin films," Thin Solid Films **727**, (2021).

[28] L.A. Walsh, C.M. Smyth, A.T. Barton, Q. Wang, Z. Che, R. Yue, J. Kim, M.J. Kim, R.M. Wallace, and C.L. Hinkle, "Interface Chemistry of Contact Metals and Ferromagnets on the Topological Insulator $Bi_2Se_3$," Journal of Physical Chemistry C **121**(42), 23551–23563 (2017).

[29] T. Manoj, Z. Wen, J. Uzuhashi, T. Ohkubo, H. Sukegawa, C. Murapaka, B. York, X. Liu, Q. Le, and S. Mitani, "Spin-Orbit Torque Modulated by Interface Chemistry in Topological BiSb/NiFe Bilayers with Titanium Insertion," ACS Appl. Electron. Mater. **6**(6), 4269–4276 (2024).

[30] S. Pal, A. Nandi, S.G. Nath, P.K. Pal, K. Sharma, S. Manna, A. Barman, and C. Mitra, "Enhancement of spin to charge conversion efficiency at the topological surface state by

inserting normal metal spacer layer in the topological insulator based heterostructure," Appl. Phys. Lett. **124**(11), 112416 (2024).

[31] X. Yao, H.T. Yi, D. Jain, M.G. Han, and S. Oh, "Spacer-Layer-Tunable Magnetism and High-Field Topological Hall Effect in Topological Insulator Heterostructures," Nano Lett. **21**(14), 5914–5919 (2021).

[32] F. Bonell, M. Goto, G. Sauthier, J.F. Sierra, A.I. Figueroa, M. V. Costache, S. Miwa, Y. Suzuki, and S.O. Valenzuela, "Control of Spin-Orbit Torques by Interface Engineering in Topological Insulator Heterostructures," Nano Lett. **20**(8), 5893–5899 (2020).

[33] L. Liu, A. Richardella, I. Garate, Y. Zhu, N. Samarth, and C.T. Chen, "Spin-polarized tunneling study of spin-momentum locking in topological insulators," Phys. Rev. B Condens. Matter Mater. Phys. **91**(23), 235437 (2015).

[34] W. Lin, K. Chen, S. Zhang, and C.L. Chien, "Enhancement of Thermally Injected Spin Current through an Antiferromagnetic Insulator," Phys. Rev. Lett. **116**(18), 186601 (2016).

[35] H. Saglam, W. Zhang, M.B. Jungfleisch, J. Sklenar, J.E. Pearson, J.B. Ketterson, and A. Hoffmann, "Spin transport through the metallic antiferromagnet FeMn," Phys. Rev. B **94**(14), 140412 (2016).

[36] S. Hikami, A.I. Larkin, and Y.O. Nagaoka, *Spin-Orbit Interaction and Magnetoresistance in the Two Dimensional Random System* (Progress Letters, 1980).

[37] S. Gautam, V. Aggarwal, B. Singh, V.P.S. Awana, R. Ganesan, and S.S. Kushvaha, "Signature of weak-antilocalization in sputtered topological insulator $Bi_2Se_3$ thin films with varying thickness," Sci. Rep. **12**(1), 9770 (2022).

[38] C. He, G. Yu, C. Grezes, J. Feng, Z. Zhao, S.A. Razavi, Q. Shao, A. Navabi, X. Li, Q.L. He, M. Li, J. Zhang, K.L. Wong, D. Wei, G. Zhang, X. Han, P.K. Amiri, and K.L. Wang, "Spin-Torque Ferromagnetic Resonance in W/Co - Fe - B/ W/Co - Fe - B/Mg O Stacks," Phys. Rev. Appl. **10**(3), 034067 (2018).

[39] H.P. Xue, R. Sun, X. Yang, J.N. Liu, Q.L. Yang, P.T. Yang, Z. Deng, X.M. Li, X.D. Bai, X.Q. Zhang, W. He, and Z.H. Cheng, "Synergistic enhancement of Gilbert damping and spin relaxation time in Fe/ $Bi_2Te_3$ heterostructures," Phys. Rev. B **111**(7), 075104 (2025).

[40] E. Saitoh, M. Ueda, H. Miyajima, and G. Tatara, "Conversion of spin current into charge current at room temperature: Inverse spin-Hall effect," Appl. Phys. Lett. **88**(18), (2006).

[41] X. Tao, Q. Liu, B. Miao, R. Yu, Z. Feng, L. Sun, B. You, J. Du, K. Chen, S. Zhang, L. Zhang, Z. Yuan, D. Wu, and H. Ding, "Self-consistent determination of spin Hall angle and spin diffusion length in Pt and Pd: The role of the interface spin loss," Science (1979). **4**, 1670 (2018).

[42] T. Manoj, V. Mattela, A.K. Salimath, S. Debroy, T.R. Kakinada, M.S. Devapriya, A. Haldar, Z. Wen, H. Sukegawa, S. Mitani, and C. Murapaka, "Thickness dependent spin to charge interconversion efficiency in polycrystalline BiSb layers deposited on Si substrate," J. Appl. Phys. **138**(5), (2025).